# IDEAL-CITIES – A Trustworthy and Sustainable Framework for Circular Smart Cities

Constantinos Marios Angelopoulos, Vasilios Katos and Theodoros Kostoulas
{mangelopoulos, vkatos, tkostoulas}
@Bournemouth.ac.uk
Bournemouth University,
Poole, UK

Andreas Miaoudakis, Nikolaos Petroulakis and George Alexandris
{miaoudak, npetro, galeksan}
@ics.forth.gr
Institute of Computer Science
Foundation for research and technology-Hellas (FORTH),
Heraklion, Greece

Giorgos Demetriou and Giuditta Morandi
{g.demetriou, g.morandi}
@pontsbschool.com
École des Ponts Business School,
Paris, France

Karolina Walędzik and Urszula Rak
{urak, kwaledzik}@bluesoft.net.pl
BlueSoft Sp. z o.o.,
Warsaw, Poland

Marios Panayiotou
m.panayiotou@cablenetcy.net
Cablenet Communication Systems
Nicosia, Cyprus

Christos Iraklis Tsatsoulis
ctsats@nodalpoint.com
Nodalpoint Systems
Athens, Greece

*Abstract*— Reflecting upon the sustainability challenges cities will be facing in the near future and the recent technological developments allowing cities to become "smart", we introduce IDEAL-CITIES; a framework aiming to provide an architecture for cyber-physical systems to deliver a data-driven Circular Economy model in a city context. In the IDEAL-CITIES ecosystem, the city's finite resources as well as citizens will form the pool of intelligent assets in order to contribute to high utilization through crowdsourcing and real-time decision making and planning. We describe two use cases as a vehicle to demonstrate how a smart city can serve the Circular Economy paradigm.

*Keywords— data-driven circular economy, IoT, crowdsensing, intelligent assets, lifelogging, citizen participation, accessibility.*

## I. INTRODUCTION

By the middle of the 21st century, the urban population living in cities will more than double, increasing from approximately 3.4 billion in 2009 to 9.8 billion in 2050 [1]. This rapid growth of urban population aggravates many challenges associated with living in urban environments, such as care services, transportation, urban pollution (noise, air, and water) etc., and particularly citizen safety, crime prevention and control [2][3][4]. The situation is exacerbated, if one considers that within the European Union, one-sixth of these citizens have disabilities and thus struggle with daily obstacles rendering them unable to participate fully in community life [5]. Cities in the 21st century fight an uphill battle to address these profound difficulties against a backdrop of shrinking environmental and financial sustainability. Tackling this complex situation requires a holistic and multidisciplinary approach towards city planning, leveraging the latest advances in technology coupled with modern social psychology and revolutionary economic thinking.

For several decades, the adopted economic model greatly focused on speed of growth and took little consideration of efficient use and management of the available resources stocks. This has led to the establishment of a *linear economy* where raw materials are extracted and then processed to produce products. These products remain within the economy with their utility decreasing over time, until they are disposed of at the end of their lifetime. Under this linear model, the production and consumption of new products directly leads to an ever increasing volume of waste, thus representing a highly inefficient use of raw material. In an effort to mitigate this, processes and methods of re-introducing waste in the economy (albeit partially) have been introduced, thus paving the way for the *recycling economy*. While this represented a qualitative extension of the linear model, recycling loops where mainly introduced towards the end of the line and their efficiency was limited as reintroduction to the economic cycle was not considered in the initial phases of product design. Consequently, the gains of recycling have also been limited as it could only be applied on a limited volume of waste. With the projected surge in consumers' demand in the near future and the stress on the natural environment now being increasingly profound, there is a need for a radically innovative economic model.

*Circular Economy* is an alternative model based on entirely different principles from the current linear economic model. Rather than using up natural resources and disposing of products when they are damaged or no longer needed, a Circular Economy maximizes the use of materials and retains their value for as long as possible (Fig. 1). A Circular Economy is based on the use of services and intelligent digital solutions, and the design and production of more durable, repairable, reusable and recyclable products. Waste is regarded as a valuable resource. Products are shared, leased or rented, rather than owned by an end user [6].

The idea of circularity comes from deep historical and philosophical roots. Particularly, to be ancient is the notion of feed-back and cycles in real-world systems, which experienced a revival in industrialised countries after World War II with the advent of computer-based studies of non-linear systems. With most recent advances, digital technology has nowadays the power to support the transition to a circular economy by drastically enhancing virtualisation, de-materialisation, transparency, and feedback-driven intelligence [7].

The Circular Economy model synthesises several major schools of thought. They include the functional service economy (performance economy) of Walter Stahel; the Cradle to Cradle design philosophy of William McDonough and Michael Braungart; biomimicry as articulated by Janine Benyus; the industrial ecology of Reid Lifset and Thomas Graedel; natural capitalism by Amory and Hunter Lovins and Paul Hawken; and the blue economy systems approach described by Gunter Pauli [8]. More recently, Kate Raworth also positioned herself among relevant contemporary authors in the field with her Doughnut Economics [9].

It is worth noting that in the modern digitized urban environment, the notions of products, assets and services are blended with those of data and information. Therefore, circularity may not refer exclusively to physical products and material but also on data that is generated as a result of the socio-economic activities of a city. This gives rise to the *Smart Circular Economy;* i.e. a data-driven economy which entails circular principles in its data management and usage as well.

## II. BACKGROUND AND RELATED WORK

### A. Policy-making on Circular Economy

Circular economy [10] is characterized as an economy that is restorative and regenerative by design, and which aims to keep products, components and materials at their highest utility and value at all times. It is conceived as a continuous positive development cycle, reforming the current economy model of 'take-make-dispose', by preserving and enhancing natural capital, optimising resource yields and minimising system risks by managing efficiently finite stocks and renewable flows.

At the beginning of 2015, the Circular Economy concept made its debut in European politics. On 2 December 2015, the European Commission published the communication "*Closing the loop – An EU Action Plan for the Circular Economy*" to help European businesses and consumers make the transition to a stronger and more Circular Economy where resources are used in a more sustainable way. The proposed actions covered the full lifecycle of products from production and consumption to waste management and the market for secondary raw materials [11]. From a European vantage point, shifting three core EU industries – mobility, food and built environment - to a circular economic process is estimated to increase productivity by 3% annually and generate EUR 1.2 trillion in non-resource and externality benefits. Compared to the current linear economic model, Circular Economy would bear total benefits of around EUR 1.8 trillion [12].

In January 2018, towards the objective of moving Europe's economy into a sustainable model and to implement the ambitious Circular Economy Action Plan, the European Commission adopted a set of measures, including a Europe-wide EU Strategy for Plastics in the Circular Economy to transform the way plastics and plastics products are designed, produced, used and recycled. By 2030, all plastics packaging should be recyclable [13]. At the same time, the Commission adopted a new proposal on Port Reception Facilities, to tackle sea-based marine litter; published a report on the impact of the use of oxo-degradable plastic on the environment [13]; and a Communication on options to address the interface between chemical, product and waste legislation that assesses how the rules on waste, products and chemicals relate to each other [14]. A Monitoring Framework on progress towards a Circular Economy at EU and national level was also implemented. It is composed of a set of ten key indicators which cover each phase – i.e. production, consumption, waste management and secondary raw materials – as well as economic aspects – investments and jobs – and innovation [15]. Finally, a Report on Critical Raw Materials and the Circular Economy that highlights the potential to make the use of the 27 critical materials in the economy more circular was also issued [16].

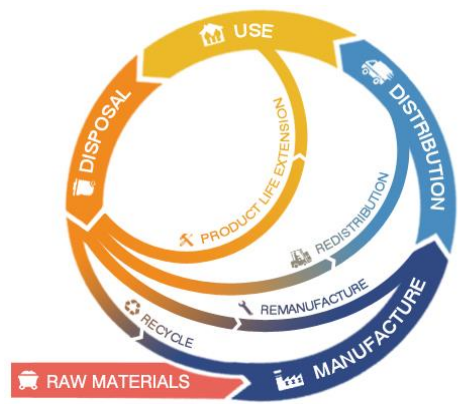

*Figure 1 The Circular Economy Cycle (https://connect.innovateuk.org/web/collaborations-circular-economy)*

Lately, on 4 March 2019, three years after adoption, the Circular Economy Action Plan was fully completed. The European Commission adopted a comprehensive report on the implementation of the 54 actions of the Plan. The report presents the main achievements and sketches out future challenges to paving the way towards a climate-neutral, Circular Economy [17].

### B. Implementation Frameworks towards Circular Economy

Moving from the regulatory to the implementation level, the shift to a Circular Economy requires innovative business models that either replace existing ones or seize new opportunities. Several are current case studies of Circular Economy implemented at a global level and many are the sectors involved in the application of circularity.

In the fast moving consumer goods sector, for example, *Braiform*, a global leader in garment hanger reuse, ensures that products are returned to reuse centres where they are sorted, repackaged and distributed back to garment-producing regions, preventing them from ending up in landfill [18]. In the electronic and electrical equipment sector, *Bundles* offers clean clothes on a pay-per-wash basis. Applying Internet of Things technology enables product monitoring, while maintenance and refurbishment of higher quality machines preserves the product integrity for multiple cycles [19].

As showed, Circular Economy finds different applications in different sectors. What is important to consider though is the added-value of new technologies applied to diverse business models. IoT, Participatory Sensing, Big Data, Cloud services, among others, enable the transition towards a model that is restorative and regenerative by design. In this context, the development of frameworks facilitating such transition has attracted research interest, particularly on the interplay and enabling synergies among ICT, Artificial Intelligence, and Circular Economy.

In [20], authors provide an initial exploration encompassing the intersection between Artificial Intelligence and the Circular Economy as the two current emerging megatrends. The paper explores the AI landscape to understand how it could accelerate the transition towards a Circular Economy at scale. Based on this approach, authors find that AI can offer substantial improvements in three main areas: product design, operations, and infrastructure optimisation.

Another working framework towards Circular Economy is provided in [21], where authors also examine the area of Circular Economy, providing an overview of the development of data driven circular approaches in manufacturing, particularly Industry 4.0, from the point of view of Reuse, Remanufacturing, Redistribution and Recycle. A key finding of this work is that research which links circular strategies and their application within Industry 4.0 to digital technologies is still a very new area of research and, as such, is an area for further studies.

The role of emerging ICT and the paradigm of Circular Economy is studied in detail in [22] via the prisms of Industry 4.0 enabling techs, and Big Data in particular. Authors note that cross-industry networks of multiple supply chains have evolved in the Circular Economy model using approaches such as industrial and urban symbiosis; however, social, environmental and economic perspectives are also highly appreciated. Therefore, data-driven operational models emerge as key components of industrial symbiosis practices.

Emerging ICT paradigms are also key enablers in transforming primary production sectors as well. In [23] Internet of Things is studied as a key enabler in transforming the metallurgical sector, while in [24] data-driven models are discussed in the context of monitoring the energy capacity of batteries for electric vehicles. A thorough review of the emergent role of digital technologies in the transition to the Circular Economy is also provided in [25].

In such innovative and promising scenario, Cities stand as one of the focal actors that can facilitate such transition. Enabling closed loops and recirculation, urban environments are hotbeds of circular economy activity, where digital technology can be used to extract and analyse data in order to create cities circular by design [24].

The IDEAL-CITIES Platform explores the role that digital technology can play in creating such urban system, particularly looking at how pairing Circular Economy and IoT provides a fertile ground for innovation and value creation. Following, we refer to related works in some key enabling ICT that are also relevant to the IDEAL-CITIES Platform.

*C. Related work on technological enablers for CE.*

*Internet of Things* (IoT) as a paradigm provisions the seamless and in large scale deployment of "smart" devices and machines. Here, "smart" refers to the ability to communicate over the Internet in order to exchange data and information both with humans and other machines. This inter-connection allows physical assets – such as sensors, actuators and objects – to be abstracted as digital resources and their operation to be optimised in the context of corresponding services.

As an example, in [26] smart use case scenarios are implemented in a smart building (e.g. adaptation of ambient conditions to human presence and preferences) by leveraging IoT and standard Internet technologies. In the context of Circular Economy, IoT and the development of corresponding smart services provides an unprecedented capability to track assets and fine-tune their usage, therefore allowing to maintain their utility at high levels.

A qualitative extension of the IoT paradigm has been underpinned in the last years by the emergence of personal smart devices (smartphones, smart gadgets, etc.) as well as of Single Board Computers - SBC (Raspberry Pi, Arduino and the likes). The former has led to an abundance of *truly personal* and *portable* devices that are well connected and fitted with a wide variety of sensors. More importantly, these are affordable devices that can be provided by the general public. SBC's are also affordable devices that have helped promote initially "at home" IoT projects and at a second stage the development of larger scale systems. The common attribute of both those technologies is the fact that needed ICT infrastructure can be provided by the general public. This has given birth to the paradigm of *IoT Participatory Sensing (IoTPS)*.

Participatory sensing was first introduced in [27], defined as the practise of employing everyday mobile devices, such as cellular phones, to form interactive, participatory sensor networks that enable public and professional users to gather, analyse and share local knowledge. *Crowdsourced Systems* [28] is a paradigm of systems that employ crowdsourcing to elicit their constituent infrastructure from the general public. For instance, in [29] an IoT system for smart buildings is presented, which opportunistically augments its sensing infrastructure to include smart devices from the occupants of the building. The wide adoption of this practice has established the notion of *crowdsensing* [30], which when applied in a Smart City context, enables the citizens to contribute to the services they consume, thus creating a network effect.

Internet of Things is emerging as one of the main sources of Big Data, providing data of high density both in space and time. Therefore, unlocking its potential to create added value requires the employment of corresponding data curation technologies, such as data analytics and machine learning algorithms, that are capable of processing data of high velocity and variety. Examples of these algorithms are those relying on online and local learning [31][32]including transfer learning and lifelong learning [33][34]. This will allow to improve the efficiency of services, production lines [35] as well as developing new business models for the Circular Economy [36]. In urban settings methodological approaches such as those of IoTPS and Crowdsensing will be of paramount importance due to the high population concentration as well as the increasing digitization of city-level services. The IDEAL-CITIES Platform leverages upon those approaches to underpin the development of trustworthy and sustainable services for Smart Circular Cities.

**Our Contribution:** The recent adoption of the Circular Economy concept, on the one hand, highlights the importance of the concepts but, on the other hand, it highlights the gap that exists both regarding the general implementation frameworks and regarding the technological enablers for this concept to be implemented. Despite the fact that some frameworks have been proposed for addressing CE, the majority of those approached the concept by focusing on a specific, rather small-scale challenges within the CE domain. Though these trials have provided some insight on specific domains, they mostly lacked a holistic approach that would allow gaining insight on the interconnection of the different entities, including the appropriate technological enablers. In the above sense, an urban environment offers an ideal environment for adopting the CE paradigm.

In this work we focus on the adoption of the Circular Economy paradigm in urban environments. First, we provide a detailed overview of recent activities on Circular Economy with respect to policy-making in the European area, implementation frameworks for Circular Economy, as well as research outcomes with a focus on key technological enablers. Then, we present the IDEAL-CITIES Platform (ICP); a novel, open and extensible platform that enables the secure and resilient acquisition and sharing of information in the context of smart circular cities. The platform leverages upon key technological enablers – such as Internet of Things, Big Data analytics, Cloud Services and Participatory Sensing – to enable the composition of services that maximize the utility of City assets (including city data) while at the same time improving the well-being and inclusivity of citizens (especially vulnerable citizen groups). The use of the platform is demonstrated via two use cases; one on assisting the movement of the visually impaired and one on increasing citizens' safety through lifelogging. The IDEAL-CITIES Platform is among the first few to demonstrate data-driven Circular Economy and to introduce the notion of circularity in smart city data.

## III. THE IDEAL-CITIES PLATFORM

In the context of the aforementioned, we present the IDEAL-CITIES Platform - ICP. ICP is designed to be a novel, open and extensible platform to enable the secure and resilient acquisition and sharing of information that is collected by individual citizens and/or authorities, through IoT on Smart Cities and participatory sensing assisted with Big Data and Cloud services. In order to account for the different data sources that can be considered in an urban environment, the low-level information, such as users' actions, would be correlated with high level one, such as context and environment information. In this sense, the IoT will feed data to the cloud services and appropriate data mining algorithms, towards accounting for the high velocity of the data arriving, as well as the heterogeneous nature of these data.

The exploitation of the aforementioned information will improve the well-being and inclusivity of citizens (especially vulnerable citizen groups), will produce more effective response to crime or other emergencies, and overall will make Smart Cities feel more secure and safe to the citizens living in them. Moreover, ICP aims to support the optimization of resource utilization and the extension of the lifecycle of IoT enabled devices. ICT will transform physical IoT-enabled assets to intelligent ones, using valuable insights and usage information based on Location, Condition and Availability (LCA) properties of such Assets.

ICP is based in the design and operational management of IoTPS applications that make use of the runtime platform in ways that can verifiably preserve required resilience, security, data integrity, confidentiality, and privacy properties, and govern the three underlying attributes enabling circularity: location, condition and availability. The key approach underpinning the development of this support will be the use of Circularity, Resilience, Security and Privacy (CRSP) patterns. CRSP patterns will provide abstract specifications of compositional structures of IoTPS applications (e.g., IoTPS service workflows, IoTPS application deployment configurations), which are proven to preserve certain composition level of the above-mentioned properties if the services/components that are composed according to the pattern satisfy other atomic level properties.

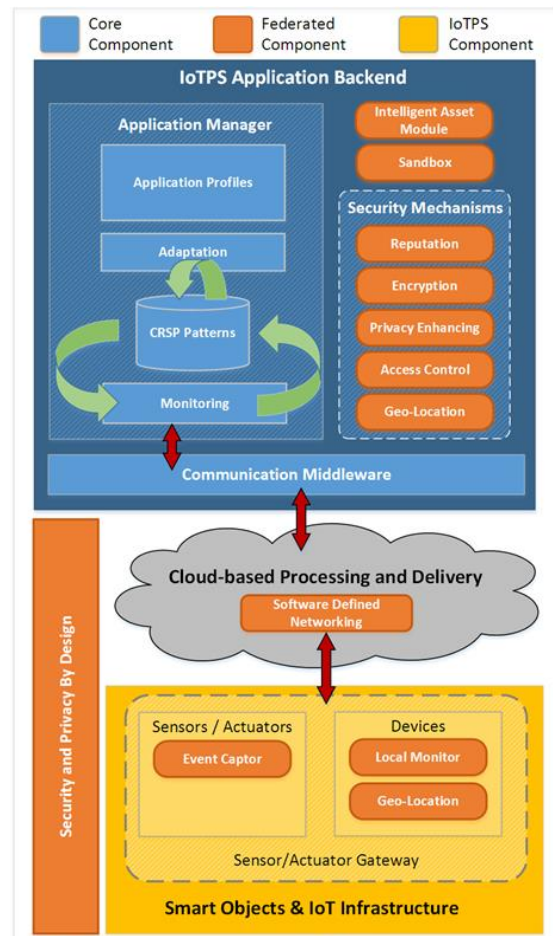

*Figure 2 The IDEAL-CITIES Platform Reference Architecture*

### A. ICP Architecture

The envisioned architecture of ICP is presented in Fig. 2. The IoTPS application backend which is a core component of ICP consists of three main parts:

a. **Communication Middleware:** Provides support for connecting applications with IoT devices that may exist in Smart Cities and/or smart devices for participatory sensing assisted with Big Data and Cloud services. The Middleware component will support the interconnection and networking of large numbers heterogeneous smart objects (devices, software, and services) based on the use of lightweight standardized communication APIs and open software components

b. **Security Mechanisms:** A set of modules providing basic device and/or user identification, authentication, access control, privacy enhancing, confidentiality maintaining, integrity and encryption functions to IoTPS applications, in order to ensure data integrity and confidentiality of private information. These will be monitored through the development of a risk register ensuring that all measures have been taken as far as reasonably practicable to maintain Confidentiality, Integrity, Availability and Resilience of data.

c. **Application Manager:** This is the main component of the IoTPS application backend, which serves as the runtime environment for the various IoTPS application profiles and contains the execution engine for the CRSP patterns.

The patterns will support both the design and operational management of IoTPS applications

B. *Advancement/Challenges*

IDEAL-CITIES approach focus on the following key aspects:

a. **IoT enabled Circular Economy business models for Smart Cities:** IDEAL-CITIES will define: (i) IoT-enabled circular business model design with focus on exploring how Cities can create new forms of value and achieve scale while transitioning to a circular business model enabled by IoT technology (exploring linkages between service and business model design), (ii) digital strategies for service business models (including public services) by analysing the role of digital resources as enablers of product-service business models.

b. **Enhance the perception of safety of the citizen to improve quality of living, better support and services:** Based on multi-disciplinary research IDEAL-CITIES will clarify which specific security mechanisms can trigger or modify the perception of citizens related to safety, security, quality of living and other support services. This will rely on: (a) the psychology of decision-making, to identify the type of bounded rationality corresponding to public roles and understand the divergence between perception and reality (b) behavioural economics, to carry out analysis of how human biases- emotional, social, and cognitive- affect security-related decisions and (c) risks, to quantify security risk (psychology of decision-making) and economic risk

c. **Confidentiality, Integrity, Availability and Resilience by Design:** IDEAL-CITIES aim is to support the design of IoTPS applications assisted with Big Data and Cloud services that can operate in a secure manner based on patterns that describe horizontal and vertical compositional structures of such applications with proven CRSP properties. The Information Asset Inventory will be able to demonstrate the level of classification for each individual data asset and how the controls required ensure user access is built in by design taking into consideration Confidentiality, Availability, Integrity and Resilience in order to comply with relevant regulations.

d. **Scalable cloud services and distributed sensors:** IDEAL-CITIES will develop algorithms for discovering, aggregating and dynamically (re-) assigning physical resources to overlay virtual entities considering the context of the data and services that request access to those entities. A key innovation will be the use of semantic information such as metadata, sensor/actuator information, sender details, network information, as well as the application requirements for describing the network resources.

e. **Geo-location in urban environments and privacy of the data exchanged:** Geo-location techniques in urban and indoor areas will be exploited in IDEAL-CITIES to provide location based context and services. Geo-location mechanisms will be developed based on the fingerprints of the wireless infrastructure of the Smart City environment without the need for additional localization hardware.

f. **Monitoring mechanisms:** IDEAL-CITIES, will develop a dynamically configurable, scalable and adaptable infrastructure that will be able to accommodate the different monitors and event captors associated with the different deployment environments, smart objects and services of IoTPS applications and will allow a scalable cross-layer IoTPS application monitoring by fusion of the different types of monitoring results generated by monitors of different layers.

g. **Decentralized information processing and the need for adaptation mechanisms:** In IDEAL-CITIES, we aim to extend the approach introduced in [37][38][39]by investigating and developing CRSP patterns describing horizontal and vertical compositional structures of IoTPS applications with proven relations between overall and component level security properties, and define possible adaptation actions for these patterns that can be dynamically adjusted based on the proven relations amongst security properties expressed in the patterns.

h. **Big Data through Participatory Sensing (PS):** IDEAL-CITIES aims to enable PS as part of applications that can also make use of IoT, Big Data and Cloud infrastructures. This requires an extension of capabilities of PS-only platforms with appropriate networking, adaptation and monitoring capabilities, a gap that will be addressed in IDEAL-CITIES

IV. ASSISTING THE MOBILITY OF THE VISUALLY IMPAIRED

Arguably, Smart Cities should not solely focus on efficiency and act as a proving ground for innovative technologies; alleviating social exclusion and the digital divide should be a core requirement and pursuit. In this context, the IDEAL-CITIES Platform will be employed to develop a use case on assisting the mobility of the visually impaired in a smart city environment. According to some estimations [40], in 2015 there were globally about 253 million people visually impaired, including 36 million blind ones.

In a high-level description, the general approach will be to aggregate and process sensory data from various sources, namely IoT-capable infrastructure and participatory sensing devices on the person to be assisted (e.g. wearables), including smartphones, in order to offer to the visually impaired persons a richer situational awareness of their surroundings; this, in turn, is expected to facilitate their navigation and movement.

Sensor-based assistive technology for visually impaired people has already a rather long history [41]. In IDEAL-CITIES, effort will be made in order to explore the possibilities made available due to recent advances in Artificial Intelligence (AI) technology for computer vision; such technology has already been made available for smartphones and other relatively low computing power devices [42] and has been constantly improving ever since [43]. Such an approach falls well within the IoTPS context (Participatory Computer Vision), comes rather naturally as a partial solution for assisting visually impaired people, and will help future-proofing the case, since such AI solutions for computer vision are expected to be ubiquitous in the near future; moreover, it can work complementarily with geo-location approaches [44]. Although the final design is still under consideration, the aim is to combine crowd-mapping with AI-based localization techniques adapted from other domains, in order to come up with working solutions for effectively assisting both the situational awareness and the movement of visually impaired persons in outdoors and/or indoors [45] environments.

## V. INCREASING CITIZEN SAFETY THROUGH LIFELOGGING

The lifelogging concept refers to the phenomenon whereby people can digitally record their own daily lives in varying amounts of detail, for a variety of purpose [46]. The multimodal data can be gathered from different sensors such as positional sensors, quantity sensor (e.g. thermometer), chemical sensors (e.g. pH of perspiration), environmental sensor (e.g. air humidity), an acoustic sensor (microphone), an optical sensor (camera e.g. private or an ambient surveillance camera), biomedical sensor (e.g. smart bracelet) or an informational sensor (e.g. smartphone collecting information about consumed or burned calories). All lifelogging devices should deliver lifelog data in a digital format necessary for further processing and storage. The lifelogging is a relatively new concept and usually refers to an ambient or passive data, but on the other hand sometimes also refers to the combination of data gained actively and passively e.g. data gathered using wearable cameras and capturing real-world information accesses [47].

Although there is a wide field of possible applications of the lifelogging phenomenon (e.g. using lifelog data for memory assistance in the Alzheimer treatment [48], [49]), the motivation of this scenario is to make use of lifelogging as an enabler for increasing citizen safety, inclusivity, security, and well-being. This will be done by empowering of the participation of citizens via usage of lifelogging application. The empowering participation approach came from the experience of the city of Barcelona claimed as the smartest city of Europe or even World. Taking a closer look at the success of the city of Barcelona [50], it can be noticed that active participation of the city stakeholders and citizens leads to effective smart cities development. Thus, it can be assumed that the usage of the lifelogging can serve as an enabler of citizens participation leading to a desired reactive intelligent infrastructure environment. Such an approach would be driven by actual citizen needs to be analysed in real-time or on-demand after a longer period of time. The overall aim of this scenario is the development of a smart secure urban environment and enabling authorities and first responders or even emergency services to be instantly notified and updated with field information depending on the level of alerts and to react on them.

The IoT perspective greatly facilities the lifelogging process to be carried out ambiently or without the lifelogger person having to initiate anything. Through lifelogging application citizens in different city areas will be able to report lifelogs passively or actively about (i) security incidents, (ii) safety incidents or (iii) life danger incidents. The effect will be evaluated in terms of quality of life, safety, and inclusivity of citizens. A wider IoT infrastructure will be able to collect and exchange different types of lifelogging citizen data, visualize it, analyse it and react by triggering different types and levels of alerts as well as take quick smart actions based on them. The proposed workflow of actions in the lifelogging scenario is presented in Fig. 3.

The example of a reported safety incident can be broken street lights (active informative lifelog) or suspicious activities in certain areas, attacks on citizens or similar incidents. At the same time, additional passive information can be gathered from body sensors. Heartbeat rate or pulse ratio, user's stress level or speed of walking can provide additional indicators of real or perceived security risks. Such information can activate other smart devices, for example, trigger alarm, increase the lighting provided by a lamp post or get help from fellow citizens who are in the vicinity of the user at risk. The reporting of life danger incidents can be done by lifelogging of health signs parameters like breath ratio, heart ratio, pulse, high temperature, level of oxygen saturation. Such lifelogs can enable quick actions in case of life danger for example during a heart attack, difficulty in breathing or loss of consciousness. Based on processing this information, additional meaningful ambient data can be collected to get more information about the accident for example via surveillance cameras connected to the same IoT infrastructure. That can help authorities to investigate citizen reports and act on them or to confirm real danger of life by the emergency services. Depending on the severity of the event other actions will be taken via city cloud like mentioned activation of the surveillance cameras, notifying emergency respondents, calling patients relatives or caregivers, contact with an In Case of Emergency (ICE) person or call to action nearby pedestrians or inform about the incident other people located nearby.

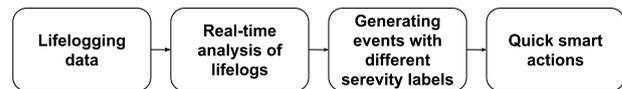

*Figure 3 Proposed workflow of taken actions in the lifelogging use case*

The data collected within the application will be analyzed in real time or storage with intentions for the further use (kept and used in accordance with the law and given permissions). Therefore, a large amount of data is to be collected in this scenario, so there are specific challenges from the Big Data area to address especially from the perspective of storage and data maintenance, organization and retrieval such as semantic access to explore search, event segmentation, visualization of lifelogs, lifelog annotation. Addressing these challenges will ease or make possible to work with acquired vast amount of data.

Summarizing, the potential outcome of the lifelogging scenario is the enhancement of citizens living environment and creating Smart Cities that are more responding to the citizen's needs, more secure, safe as well as providing a more effective response to the crime and other emergencies like citizen health danger. The improvement of well-being and inclusivity of the citizens will be a consequence of these changes.

## VI. CONCLUSION

Throughout the history of human evolution, the technological advancements have in principle contributed to the improvement of the quality of life. However, each disruptive technological innovation has had its toll at the planet's finite resources, as the exploitation of these typically was based on the linear economy approach.

The introduction of ICT technologies in a city's core infrastructure is an opportunity to employ the underlying technological means to amortise the adverse effects of the linear economy. Data-driven CE is the intersection of ICT and Circular Economy and a promising way forward to deliver CE in a large scale. In this paper we introduced a reference architecture for a smart circular city and two associated use cases. These particular use cases are representative and illustrate the depth and breadth of the challenges future cities will face, such as inclusivity, security and privacy, which essentially relates to citizen's acceptance. However, despite these use cases are informative and expand our understanding on the circular aspects of the economy, certain limitations are imposed, mainly due to the fact that both cases would have to address specific scenarios and limited IoT devices. In this sense, though there will be a chance to uncover previously unseen dimensions and limitations within the Circular Economy concept, further pursuing of these dimensions would be required.

ACKNOWLEDGMENTS

This work has been partially supported by IDEAL-CITIES; a European Union's Horizon 2020 research and innovation staff exchange programme (RISE) under the Marie Skłodowska-Curie grant agreement No 778229.